# The signature of supernova ejecta measured in the X-ray afterglow of the Gamma Ray Burst 011211

J.N.Reeves*, D.Watson*, J.P.Osborne*, K.A. Pounds*, P.T. O'Brien*, A.D.T. Short*, M.J.L. Turner*, M.G. Watson*, K.O.Mason†, M.Ehle‡, N Schartel‡

*X-ray Astronomy Group, Department of Physics and Astronomy, University of Leicester, University Road, Leicester LE1 7RH, UK

† Mullard Space Science Laboratory, University College London, Holmbury St. Mary, Dorking, RH5, 6NT, UK.

‡ XMM-Newton SOC, Villafranca, 28080, Madrid, Spain

**Since their identification with cosmological distances[1], Gamma-ray bursts (GRBs) have been recognised as the most energetic phenomena in the Universe, with an isotropic burst energy as high as $10^{54}$ ergs[2]. However, the progenitors responsible for the bursts remain elusive, favoured models ranging from a neutron star binary merger[3-5], to the collapse of a massive star[6-8]. Crucial to our understanding of the origins of GRBs is the study of the afterglow emission, where spectroscopy can reveal details of the environment of the burst. Here we report on an XMM-Newton observation of the X-ray afterglow of GRB 011211. The X-ray spectrum reveals evidence for emission lines of Magnesium, Silicon, Sulphur, Argon, Calcium, and possibly Nickel, arising in enriched material with an outflow velocity of order 0.1c. This is the first direct measurement of outflowing matter in a gamma ray burst. The observations strongly favour models[30] where a supernova explosion from a massive stellar progenitor precedes the burst event and is responsible for the outflowing matter.**

The Gamma Ray Burst GRB 011211 was first detected on December 11$^{th}$ 2001 at 19:09:21 (UT), by Beppo-SAX[9]; the burst duration was 270s (making GRB 011211 the longest burst observed by Beppo-SAX) with a peak flux (40-700 keV) of 5×10$^{-8}$ erg cm$^{-2}$ s$^{-1}$. Spectroscopy of the optical afterglow revealed several absorption lines at a redshift of z=2.141±0.001[10,11] and R band imaging[12] has linked the optical transient with extended emission - the probable host galaxy - of magnitude m$_v$=25.0±0.5. Assuming the absorption system arises from the GRB host galaxy and adopting a cosmology of H$_0$ = 75 km s$^{-1}$ Mpc$^{-1}$ and q$_0$=0.1 implies a total equivalent isotropic energy for GRB 011211 of 5×10$^{52}$ erg.

The XMM-Newton[13] observations of GRB 011211 started at 06:16:56 (UT) on December 12$^{th}$ 2001, 11 hours after the initial burst[14]. Data from the European Photon Imaging Cameras (EPIC) have been analysed, using both the MOS and PN instruments; the total observation duration is 27 ks. The time-averaged 0.2-10 keV flux was 1.9x10$^{-13}$ erg cm$^{-2}$ s$^{-1}$ (corresponding to a 0.6-30 keV rest frame afterglow luminosity of 7×10$^{45}$ erg s$^{-1}$), decreasing during the observation as F(t) ∝ t$^{-(1.7\pm0.2)}$. The Optical Monitor detected the burst afterglow in both the visible and ultra-violet (UVW1) bands (at 11 and 6σ confidence), with magnitudes of V=21.12±0.13 and UVW1=21.6±0.3. The magnitudes correspond to fluxes of (1.28±0.16)×10$^{-17}$ erg cm$^{-2}$ s$^{-1}$ Angstrom$^{-1}$ and (0.85±0.23)×10$^{-17}$ erg cm$^{-2}$ s$^{-1}$ Angstrom$^{-1}$ at wavelengths of 5430 Angstrom and 3140 Angstrom respectively. These data are consistent with reddening in the host galaxy of E(B-V)~0.2, or with Lyman alpha absorption at high redshift. Figure 1 shows the time-averaged EPIC X-ray spectrum of the burst afterglow. The continuum can be fitted by a power-law of photon index Γ=2.3±0.1, attenuated by the Galactic absorption column of 4.2×10$^{20}$cm$^{-2}$. An apparent absorption feature is present between 0.3 and 0.4 keV, which when fitted by an absorption edge, gives an energy of E$_0$=0.99±0.09 keV and an optical depth of τ=1.0±0.4, in the burst rest frame.

Spectra were initially extracted in five time segments, corresponding to 0-5 ks, 5-10 ks, 10-15 ks, 15-20 ks and 20-27 ks after the start of the XMM-Newton observation. As significant line emission was observed only during the initial 10 ks, the remaining data was binned into one 17 ks time bin. Figure 2 shows the EPIC-pn spectrum between 0.2 and 3 keV, from the first 5 ks only; emission lines are detected in the burst rest-frame at energies 1.40±0.05 keV, 2.19±0.04 keV, 2.81±0.04 keV, 3.79±0.07 keV and 4.51±0.12 keV. The closest abundant K$\alpha$ transitions to the observed lines are: - Mg XI (1.35 keV) or Mg XII (1.47 keV), Si XIV (2.00 keV), S XVI (2.62 keV), Ar XVIII (3.32 keV) and Ca XX (4.10 keV). Thus we infer that the lines are *blue-shifted* with respect to the known redshift of GRB 011211. To quantify this, we adopted the rest-frame energies corresponding to Mg XI, Si XIV, S XVI, Ar XVIII and Ca XX and varied the redshift of the line emitting material. The best-fit redshift for the line set was found to be $z=1.88\pm0.06$, differing significantly (at >99.99% confidence) from the known GRB redshift of $z=2.140\pm0.001$, implying an outflow velocity for the line emitting material of $v/c=0.086\pm0.004$, or $v=25800\pm1200$ km s$^{-1}$. The line emission also declines more rapidly than the continuum (at >3$\sigma$ confidence), suggesting a more enduring (non-thermal) component to the continuum flux. The decrease in flux of the Si XIV line is shown in Figure 3, the line is detected only over the first 10 ks of observation.

In order to assess the quality of the spectral fit and the statistical significance of the emission lines, we calculated the fit statistic, measured in terms of total $\chi^2$ deviations between the data points and the input model, divided by the degrees of freedom (d.o.f) in the fit. This improved from $\chi^2$/d.o.f=56.7/47 for a pure power-law model to 36.2/41 upon the addition of the lines. Employing an F-test[15] then yields a significance level of 99.7% for the set of lines as a whole. Furthermore, by performing Monte-Carlo simulations, we find a probability of only 0.02% that the lines result from purely random Poisson noise. We conclude that the line emission is detected with good confidence.

X-ray line emission can arise by a variety of processes, including thermal emission, recombination in a photoionised plasma, or by reflection of hard X-rays in dense, optically thick matter. Modelling the early spectrum with emission from an optically thin plasma of temperature kT=4.5±0.5 keV and luminosity $7\times10^{45}$ erg s$^{-1}$ (using the 'MEKAL' code[16]), requires an over-abundance for Mg, Si, S, Ar and Ca of ~10 times the solar value (solar abundance is ruled out at >99.9% confidence). The absence of emission in the Fe K band gives a limit to the abundance of iron at <1.3 solar. The fit statistic for the thermal model is excellent ($\chi^2$/d.o.f=37/44), Monte-Carlo simulations also showed that only one in ten thousand pure power-law spectra can yield as good a fit by random chance alone (i.e. a null probability of ~0.01%). Another possible source of the line emission is X-ray reflection[17,18]. This is currently favoured in 'nearby reprocessor' models[19], where X-rays scatter off dense material within the stellar envelope of a massive progenitor star. Ionised reflection models[17] were fitted to the data for two cases: - (i) where the walls of the stellar envelope subtend $2\pi$ steradian solid angle to the X-ray source; (ii) where the emission arises purely from the scattered X-ray flux, with no continuum emission. In both cases the fit obtained is poor ($\chi^2$/d.o.f = 62.4/46 and 56.7/47 respectively), the reflection models fail to reproduce the emission lines observed in the early afterglow spectrum.

A major drawback of both the photoionisation and reflection models is that they favour emission from heavier metals such as Fe or Ni, rather than the lighter elements (since the line fluxes increase as the atomic number squared), which is difficult to reconcile with the observed line spectrum. Therefore we consider the thermal model to be physically plausible, noting that the non-detection of Fe line emission is then a less severe constraint. We assume that a dense shell of material (the ejecta of a very recent supernova) is heated by the gamma ray burst. At the densities deduced below, the emitting plasma will rapidly reach thermal equilibrium and will radiate over a timescale given by $t_{cool} \sim 2\times10^{15}\, n_e^{-1}\, T_8^{1/2}$ s, where $T_8$ is the plasma temperature in units of $10^8$ K, and $n_e$ is the electron density (in $cm^{-3}$). From the observed X-ray luminosity, we estimate an emission measure $n_e^2 V = 10^{69}$ $cm^{-3}$, where V is the total emitting volume, whilst the temperature is also derived from the thermal spectrum (4.5 keV or $5 \times 10^7$ K). We interpret the observed emission line timescale (t) in terms of the light travel delay in radiation arising from different parts of the illuminated shell. For a spherical shell this delay depends on the radius of the shell R and the beam half-opening angle θ, and is given by $R=ct/(1-\cos\theta)$. Converting the duration of the line emission to the burst rest-frame, $t=10^4/(1+z)$ sec, for a half opening angle of θ~20° (consistent with GRB beaming models[20]), yields R~$10^{15}$cm. Combining the emission measure of illuminated matter, with an electron scattering optical depth ~1 for the shell, then allows estimates, for R~$10^{15}$cm, of the electron density $n_e$~$10^{15}$ $cm^{-3}$ and shell thickness ~$10^9$ cm (in practice the ejecta may be clumpy and thus distributed over a thicker shell). The mass of the illuminated matter is then approximately 0.1 solar masses (~$10^{32}$ g) with a kinetic energy of ~$10^{51}$ erg, consistent with the energies of a typical supernova and the gamma ray burst (if beamed). In the isotropic limit, where the thermal emission results from spherical shell of solid angle 4π steradian, then the radius of the shell will simply be R=ct, yielding R~$10^{14}$ cm; this sets a minimum distance between the burst and the line emitting material. One may also expect to see redshifted line emission from the counter-jet receding from us, however matter outside the beam may prevent the X-ray emission from the far side of the shell being seen.

With a mean outflow velocity of the ejecta of 0.1c, the estimated radius of the shell, R~ $10^{15}$ cm implies a time delay between the supernova and gamma ray burst of ~4 days (in the isotropic limit, this time delay will be a minimum of 10 hours). This offers a natural explanation for the non-detection of Fe K emission in the XMM-Newton spectrum. The stable isotope $^{56}$Fe is produced by β-decay via the reaction $^{56}$Ni → $^{56}$Co → $^{56}$Fe, the reactions having half-lives of 6 and 78 days. Over a short timescale after the supernova, as our model implies, very little Fe will have been synthesised. Instead we would expect to observe spectral features due to Ni, or Co. Interestingly we do find a marginal detection (90% confidence) of the blueshifted Ni K$\alpha$ emission line, with a rest frame equivalent width of 800 eV (equivalent to an eventual iron abundance of 4 × solar). While iron will be present in the collapsed core of the massive progenitor, most of the illuminated ejecta will be from the outer stellar layers (i.e. the lower Z elements). Our result contrasts with the large masses of iron implied by the iron lines reported in the X-ray spectra of several other afterglows[21-23]. It seems plausible that in some of those cases the actual detection was of Ni K$\alpha$[24]; otherwise a much longer delay between an initial supernova and the onset of the gamma ray burst is required, of the order several months.

Importantly, the XMM-Newton observation appears to rule out a neutron star binary merger as the progenitor of GRB 011211, as such a merger is unlikely to expel sufficient quantities of matter into the surrounding medium[25], nor can the relatively low iron abundance be explained, as any supernova will have occurred many years before the stellar merger. The combination of a high density, over-abundant light metals, and a high velocity outflow in the X-ray emitting plasma strongly suggests an association of GRB 011211 with a very recent supernova, caused by the collapse of a massive progenitor star[26]. In this scenario the supernova ejects a substantial quantity of enriched material at high velocity (~0.1c) into the surrounding medium, which is subsequently illuminated by the gamma ray burst. The high metal abundances of the illuminated material will arise naturally from the pre-supernova nucleosynthesis and the short interval between the supernova explosion and the burst itself. Indeed the supernova model is theoretically preferred over neutron star mergers to explain long-duration gamma ray bursts[7], as seen here. The supernova link is further strengthened by the observed correlation of long-duration bursts with star forming regions[27,28] and in one case a direct association between the bright SN 1998bw and the very low redshift (z=0.0086) burst, GRB980425[29].

Acknowledgements

This paper is based on observations obtained with XMM-Newton, an ESA science mission with instruments and contributions directly funded by ESA Member States and the USA (NASA). We are grateful to Sir Martin Rees, Andrew King, Melvyn Davies, Brian McBreen, Dick Willingale, and Martin Barstow for critical reading of the paper and discussions.

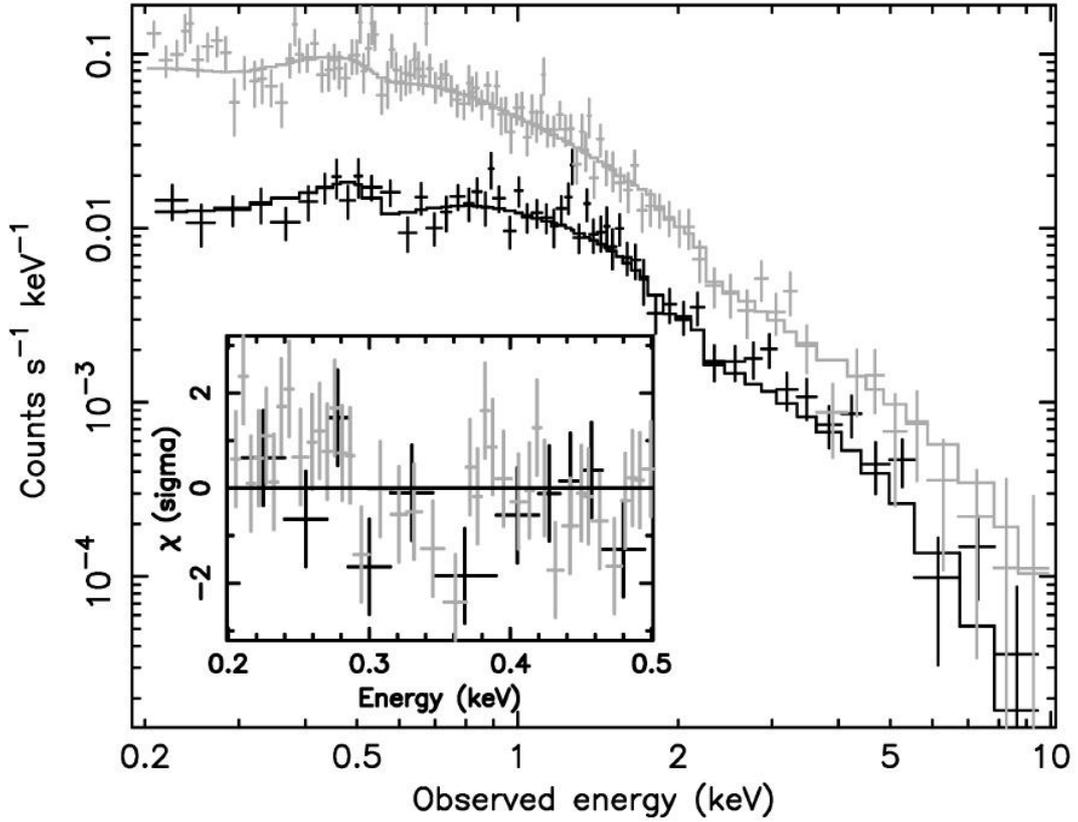

Figure 1: - The XMM-Newton spectrum of the afterglow of gamma ray burst, GRB 011211. The spectrum shown is integrated over the whole 27 ksec observation. EPIC-pn detector data points are shown in grayscale, EPIC-MOS detector points in black. The ordinate shows the counts per second obtained in each detector, over a given energy range (keV), the abscissa plots the observed energy (in the observed frame) of the X-ray photons. The solid curves show the model fit, convolved through the instrument response and energy resolution. In this case, the continuum emission can be modeled by either a power-law of photon index $\Gamma=2.3\pm0.1$ or by thermal bremsstrahlung emission of temperature $kT=2.1\pm0.2$ keV, attenuated only by an absorption column of $4.2\times10^{20}$cm$^{-2}$ from our own galaxy. An apparent absorption feature is present between 0.3 and 0.4 keV (see inset), this can be modeled by an absorption edge (where the opacity after the edge energy $E_0$ falls as $E^{-3}$). In the burst rest frame the edge energy is $E_0=0.99\pm0.09$ keV with an optical depth $\tau=1.0\pm0.4$.

Assuming the above absorption feature arises in the same material as the line emission, then its likely identification is with the OVIII edge at 0.87 keV. Note however that although the absorption edge is formally significant (99.9% from an F-test), the detector resolution is lowest at this energy ($\Delta E \sim 100$ eV at 300 eV), which makes the shape of the feature difficult to constrain.

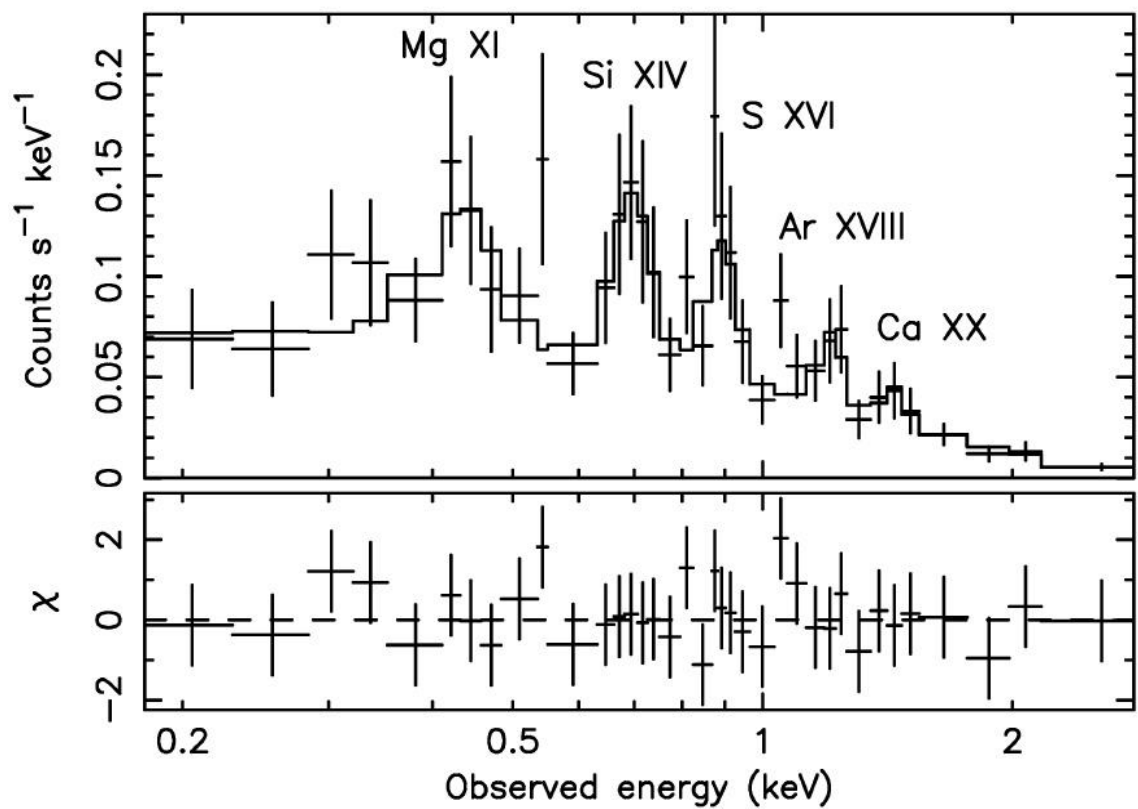

Figure 2: - The XMM-Newton EPIC-pn spectrum of the burst afterglow, for the first 5 ks of exposure only. The upper panel shows the observed count rate spectrum, whilst the lower panel plots the residuals of the thermal model compared with the data points, in units of 1 $\sigma$ deviations. The energy plotted on the abscissa is in the observer frame. Emission lines are detected at energies of 0.45±0.03 keV, 0.70±0.02 keV, 0.89±0.01 keV, 1.21±0.02 keV and 1.44±0.04 in the observed spectrum, whilst the measured line fluxes are $(7.6\pm5.1)\times10^{-15}$

erg cm$^{-2}$ s$^{-1}$, (1.1±0.3)×10$^{-14}$ erg cm$^{-2}$ s$^{-1}$, (9.9±2.9)×10$^{-15}$ erg cm$^{-2}$ s$^{-1}$, (6.7±2.5)×10$^{-15}$ erg cm$^{-2}$ s$^{-1}$ and (4.4±2.2)×10$^{-15}$ erg cm$^{-2}$ s$^{-1}$ respectively. These correspond to energies of 1.40±0.05 keV, 2.19±0.04 keV, 2.81±0.04 keV, 3.79±0.07 keV and 4.51±0.12 keV in the burst rest frame (at z=2.14), with rest-frame equivalent widths of 180±120 eV, 430±130 eV, 480±140 eV, 460±170 eV and 360±180 eV respectively (all errors are quoted at 1σ confidence). Note that an iron K line is not detected; the upper-limit on the rest-frame equivalent width is <400 eV. For reference, the Full Width Half Maximum resolution of the EPIC-pn spectrum is 100 eV at 1 keV. The emission lines can be identified with the Kα transitions of Mg XI (or Mg XII), Si XIV, S XVI, Ar XVIII, and Ca XX. The observed energies are blue-shifted by a factor corresponding to a velocity of v=0.086c (or 25800 km s$^{-1}$), when compared to the redshift of the gamma ray burst (at z=2.14). For clarity only the EPIC-pn data are shown; consistent results are obtained for the EPIC-MOS camera, although the signal to noise of the MOS data is lower.

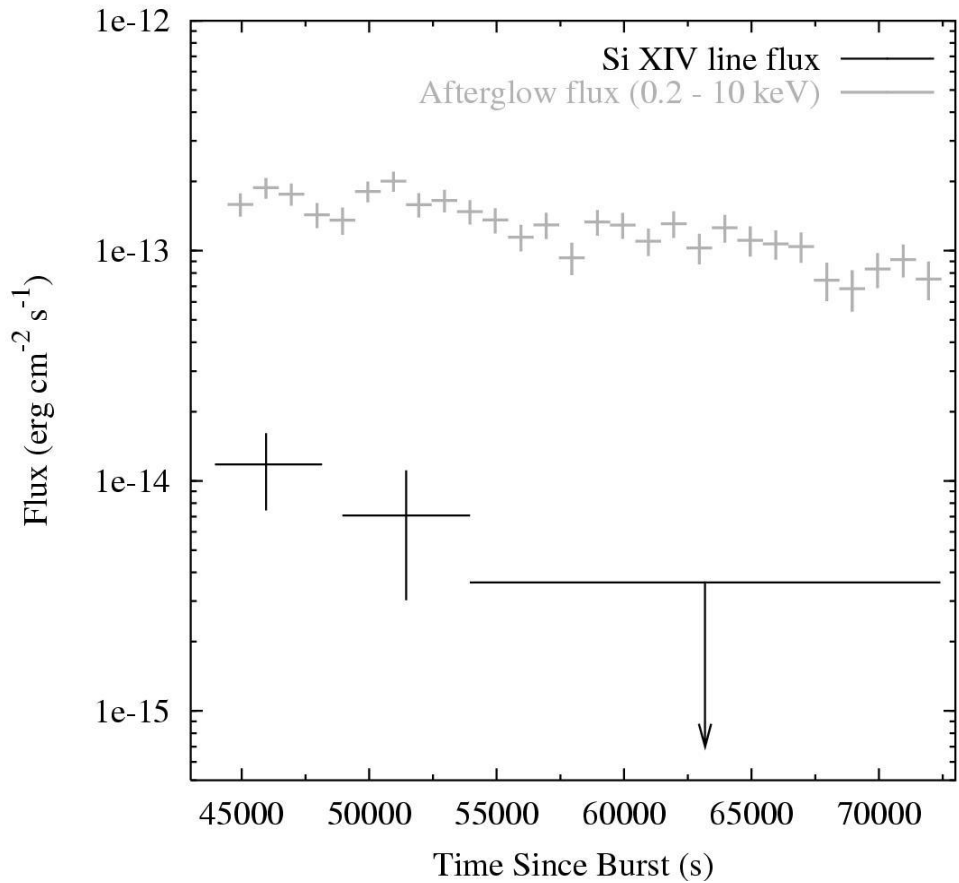

Figure 3: - The line flux of Si XIV Kα, plotted as a function of time since the initial burst. Also plotted is the total continuum flux (0.2-10 keV) against time. The Si XIV line is detected in the first 10 ks of the observation (illustrated by the first two data points), but is not during the remainder of the observation (the upper-limit shown is at the 3σ confidence level). The values of the Si XIV line flux are $(1.1\pm0.3)\times10^{-14}$ ergs cm$^{-2}$ s$^{-1}$ during the first 5 ksec, $(0.7\pm0.4)\times10^{-14}$ erg cm$^{-2}$ s$^{-1}$ from 5-10 ksec and $<3.6\times10^{-15}$ erg cm$^{-2}$ s$^{-1}$ during the last 17 ksec of observation. A similar effect is seen for the other emission lines, none being detected in the later part of the observation. For instance, the S XVI line flux is $(1.0\pm0.3)\times10^{-14}$ erg cm$^{-2}$ s$^{-1}$ in the first 5 ks, decreasing to $<2.5\times10^{-15}$ erg cm$^{-2}$ s$^{-1}$ during the last 17 ks of observation.